*The talk for the 1995 World Congress on Neural Networks*

# MultiNeuron - Neural Networks Simulator For Medical, Physiological, and Psychological Applications

A.N.Gorban[1*], D.A.Rossiyev[2], M.G.Dorrer[3]

[1]Computing Center Russian Academy of Sciences,
[2]Krasnoyarsk State Medical Academy
[3]Krasnoyarsk State Technical Academy

## ABSTRACT

This work describes neural software applied in medicine and physiology to:
- investigate and diagnose immune deficiencies; diagnose and study allergic and pseudoallergic reactions; forecast emergence or aggravation of stagnant cardiac insufficiency in patients with cardiac rhythm disorders; forecast development of cardiac arrhythmia after myocardial infarction; reveal relationships between the accumulated radiation dose and a set of immunological, hormonal, and biochemical parameters of human blood and find a method to be able to judge by these parameters the dose value; propose a technique for early diagnosis of choroid melanomas;

Neural networks help predict human relations within a group.

## MULTINEURON SOFTWARE

MultiNeuron software is a base manager for neural networks making possible to create and train neural networks on the basis of various training samplings, to store the networks on computer discs and test examples. The program generates and governs 3 types of neural networks designed to solve different problems:

The program features three training modes:
1. basic;
2. excluding examples "standing out" of the assigned training model. This mode makes possible to correct the assigned training model and point out the subgroups of examples which are separate training samplings;
3. minimization of training parameters, making possible to find a minimal set of them required for successful training and testing.

Reliability of training neural networks can be different; like other parameters, it can be altered in the course of training.

Main algorithms of training and operation are described in monograph [1]. Applications of learning neural networks can be found in works [2-6].

Use is made of full connection neural networks: having received the input signals the neurons exchange between themselves signals a certain number of times; after that the output signals are taken from them. The first neural simulators of this series have been described in detail in [1].

---

[*] Present address: University of Leicester, Leicester LE1 7RH, UK,
e-mail: ag153@le.ac.uk

**Parameter significance determination.** In the course of training for a neural network to make a decision about the answer is frequently required to estimate relative significance of each training parameter. Estimation of training parameters' significance in MultiNeuron program makes possible to find automatically their minimal set in every training sampling. The most vivid case is reduction of the number of parameters necessary to forecast emergence or aggravation of cardiac insufficiency symptoms in patients with various cardiac rhythm disorders from 97 to 4. Below is a brief description of MultiNeuron program applied in medicine, psychology and physiology.

## PSYCHOLOGICAL INTUITION OF NEURAL NETWORKS

In this work we employ trainable neural networks to start solving two problems facing the designers and users of computer psychological tests:
1. Cultural, national and social adaptation of tests.
2. Gaining direct recommendations and forecasts avoiding the phase of explicit diagnosing (constructing a "measured individuality"). If the explicit diagnoses (every form of verbal typological description) are not used, the recommendations and forecasts are taken as intuitive. In this sense we speak about the psychological intuition of the neural networks.

## EXPERIMENT: PREDICTION OF GROUP RELATIONS AVOIDING EXPLICIT DESCRIPTIONS

We put the question of feasibility to apply neural networks to determine relations within a group, and feasibility to give recommendations to correct the patient's condition avoiding construction of a "described individuality".

The experiment was supposed to confirm (or reject) the hypothesis that a neural network makes possible to simulate relations within a group of people on the basis of their psychological properties and yield a forecast for a new person to become a member of the group.

For initial data we used results of polls made in parallel in a student academic group.

The input data formed from the results of the poll of persons under investigation by the psychological questionnaire of 91 question grouped in 3 groups; the purpose of the questionnaire was to define constant psychological properties of a person. Information was coded by "yes" =1, "don't know-0, "no"=-1 principle. The questionnaire was compiled on basis of standard techniques (MMPI and other).

Special attention should be paid to the number of questions - 91. Standard techniques of MMPI type have hundreds of questions. The answers of a person to these questions contain hundreds of bits of information. To make diagnosis reliable ("measurements of an individuality") these hundreds of bits are compacted to a score of bits (more rarely - to several scores). With the classical approach this score of bits is the "result of the test". Later this forms the basis to make prognoses and recommendations. We suggest to use directly the entire initial database to train the neural networks and to use them further. It turns out that such a direct use makes possible to reduce the questionnaires, leaving several scores of answers only (in this experiment - 91, further reduction is possible).

To form the initial data we used the results of socio-metric study of an student academic group by the question: :"To what degree would you like to work in your future profession with each group member?". The answer was supposed to be given as a 10-point estimate (0 - most negative attitude to a person as a would-be co-worker, 10 - maximum positive). After the questionnaires with these answers were collected statistical characteristics of the socio-metric matrix were calculated - an estimate of the status and expansivity of each group member. These characteristics were used as elements of neural network output vector.

To solve the problem we used the method of council of networks with different characteristics (it was presumed to obtain at this comparative characteristics of networks with different neuron number).

Members of this council were 6 neural networks with different characteristics, (the 1st, the 2nd and the 5th with 64 neurons, the 3rd, the 4th and the 6th - with 32 neurons).

Maximum error in the answer was less than 23%, which, specificity of the subject under investigation considered, is an excellent result. Mean square error estimate yield even more impressive results - it doe not exceed 9.4% (in socio-metric studies 10% is considered to be a good results and standard for study reproducibility).

Thus, neural networks can be used to simulate relationships in a human group and forecast relations with the group of a new member. This is possible without formulating an explicit diagnosis and constructing measured individuality. What's more, abandoning the explicit diagnosis leads to economy in the most labor-intensive phase - obtaining the initial information (91 question instead of several hundreds!).

The results obtained encourage: the psychological intuition of the neural networks can form the basis to construct a new generation of methods for psychological forecast and consulting (recommendations). Abandoning the intermediate phase of explicit diagnosis makes possible to reduce the quantity of required initial information retaining the reliability.

## NEURAL NETWORKS IN INVESTIGATION AND DIAGNOSIS OF IMMUNE DEFICIENCIES

Subject to medical examination were 20-25 years old people both healthy (51) and with the clinical diagnosis of immune deficiency (ID) (42). Activity of basic lymphocyte intracellular enzymes was studied, cell immunity was analyzed. Patients were diagnosed as ID on the basis of clinical examination. The input data in the course of training the neural network was a training sample consisting of examples each being a set of laboratory parameters of one person. The diagnosis was the answer to an example (in this case - sick or healthy). The training sampling consisted of examples with both the former and the latter answers known. The neural networks were trained to discern healthy and sick people by 17 metabolic parameters of lymphocytes. In the course of training the neural network used all examples of healthy persons, but from the group of sick persons it excluded 30 people not fitting the preassigned classification model. The remaining 12 persons were used by the network for training. Statistical analysis of the subgroup obtained revealed that basing on lymphocyte enzyme activity indices the neural network divided the entire sampling of persons with ID into two groups different in exchange of substances in lymphocytes and in the type of immune response. Seemingly, the ID of the persons remaining and excluded are at different stages. The remaining group given the depressed immune response betrayed the first stage of compensation reaction of the immune system, revealing, primarily, in higher activity of intracellular enzymes. In the excluded persons most metabolic indices approach those of healthy people and they, probably, betray the second stage of the compensation reaction manifested in activated immune reactivity. After the training sampling was added immune status parameters and 3-group classification model was assigned the neural network successfully learned to discern healthy people from sick patients and determine the ID stage thus confirming the optimality of the given classification model.

## DIAGNOSIS OF ALLERGIC AND PSEUDOALLERGIC REACTIONS

Allergic reactions are frequently not easy to diagnose. Pseudoallergic reactions clinically similar to allergic but distinct from them immunologically add to the complexity of the problem. The pseudoallergic reaction are frequent among patient with drug and food intolerance and feature certain peculiarities in pathogeny, course of disease and its treatment. The aim of investigation was to apply the neural network classifier to create an automated diagnosticum for allergic and pseudoallergic reactions as well as to study pathophysiological peculiarities of these reactions. For the input parameters we used activity levels of NAD(P)-dependent dehydrogenases and concentration of certain metabolites in blood lymphocytes. The initial classification model had 3 classes: 1 class - healthy people, 2 - really allergic patients, 3 - pseudoallergic patients. The training of a neural network with few neurons revealed that to make training successful the classification should be increased - the 4th class is distinguished. It comprises the people from all three classes difficult to diagnose on the basis of the given set of parameters (about 17% of people screened). The persons of the 4th class have to be further examined more thoroughly.

Most significant among the input parameters were the intracellular substrate levels and cofactors. Analysis of significance of the input parameters gives a new view at the distinctions in pathogeny between the allergic and pseudoallergic reactions.

## FORECAST OF EMERGENCE OR AGGRAVATION OF STAGNANT CARDIAC INSUFFICIENCY IN PATIENTS WITH CARDIAC RHYTHM DISORDERS

Therapeutics Department N 1 of Krasnoyarsk State Medical Academy examined 70 patients with various cardiac rhythm disorders. All patients were examined clinically, underwent electrocardiography, veloergometry, echocardiography, transesophageal electric stimulation, etc. Later frequent faints due to the basic pathology made implant 20 of them electric cardiac stimulators (ECS). All patients were re-examined in 1-10 years; this examination made possible to divide them into 4 groups:
1. with no ECS implants without changes in the stage of cardiac insufficiency (CI);
2. with no ECS implants who emerged or aggravated CI symptoms;
3. with ECS implants without changes in CI stage;
4. with ECS implants who emerged or aggravated CI symptoms.

The neural networks were trained to forecast emergence or aggravation of CI symptoms depending on ECS expected. For this two neural networks were used, one of them was trained to discern groups 1 and 2, the other - 3 and 4. In this fashion the neural networks may be able to forecast aggravation or emergence of CI symptoms by the data of the primary examination of patients with different cardiac rhythm disorders. To validate the prognostication capabilities of neural networks all patients have undergone testing by both neural networks. The testing data showed the ECS implantation in group 2 patients to reduce considerably the prognosticated probability of developing or aggravating cardiac insufficiency. Modeling of ECS absence in group 3 patients either does not change the forecast or considerably aggravates it depending on the type of pathology. The research conducted makes possible to simulate presence of ECS in patients with cardiac rhythm disorders make decisions about its implantation.

## EXPERT SYSTEM TO FORECAST ONE OF MYOCARDIAL INFARCTION COMPLICATIONS (CARDIAC ARRHYTHMIA)

Neural classifiers used to design an expert system to forecast one of myocardial infarction complications (cardiac arrhythmia) estimated the significance of input parameters. The input of each of 4 neural network experts was fed 36 parameters of a patient after an infarc-tion: patient's sex, age, cardiac muscle site affected, some electrocardiogram and echocardiogram parameters, clinical symptoms upon arrival at the clinic, medicinal preparations prescribed.

In the opinion of neural network experts most significant parameters important for predicting the complication are the heart rate recorded by the electrocardiogram, hypertension, sodium content in blood, disturbances of the cardiac rhythm in patient's medical history. When modeling the effect of these parameters on the forecast the neural network helped find out that their effect on the risk of cardiac arrhythmia development is different: heart rate increase, for example, increases the risk of development of the permanent form of cardiac arrhythmia and decreases the risk of development of paroxysmal (transient) form. Presence and phase of the hypertension render quite an opposite effect. Thus, in addition to practical application in clinical practice this neural network expert system can be a tool for scientific research to study pathophysiological mechanisms of the disease.

## RELATIONSHIP BETWEEN THE ACCUMULATED RADIATION DOSE AND A SET OF IMMUNOLOGICAL, HORMONAL, AND BIOCHEMICAL PARAMETERS OF HUMAN BLOOD

Our work was aimed to find relationship between the accumulated radiation dose and a set of immunological, hormonal, and biochemical parameters of human blood and find a method to be able to judge by these parameters the dose value. To this end we used the neural network classifier. The training examples were the above said parameters (35) in 141 persons working at a multi-production enterprise of nuclear industry. The people under examination have undergone radiation monitoring of the dose accumulated and divided into 3 classes depending on the dose magnitude. Class 4 were people with the dose within the limits of natural background values.

Statistical analysis of the training sampling most parameters did not reveal reliable distinctions between the classes, while those available could not be traced over all classes. However, using the whole set of parameters the neural network successfully learned to recognize the class of each example. When testing the sampling with the answers known the class was determined correctly in 100% cases. The trained net tested 140 people living in the area of the nuclear industry enterprise who are not engaged in production. By the data of the neural network only 3 persons had the accumulated dose within the limits of the natural background; 51 person betrayed a weak dose, 82 - medium, 4 persons - high. Neural technology for early diagnosis of malignant tumors in vascular coat of the eye. Melanomas of the vascular coat of the eye account for 88% of all intraocular tumors, yet they are easily diagnosed at the last stages of disease development only. The available methods to reveal the disease at its early stage are not very much reliable and rather complicated and costly.

## EARLY DIAGNOSIS OF CHOROID MELANOMAS

The proposed technique for early diagnosis of choroid melanomas involves a new laboratory method developed at Krasnoyarsk P.G. Makarov Ophthalmologic Center and interpretation of the obtained data by the neural network classifier.

The laboratory method is based on indirect measurement of pigment (melanin) content in eyelashes. The data of spectrophotometry done at several frequencies for each eye and some general characteristics of the patient (sex, age, etc.) are fed onto the input synapses of a 43-neuron classifier with associative input matrix. The neural network decides whether the patient has a tumor and in the event of a positive answer determines its stage. with confidence in per cent. Even a slight suspicion of a tumor may be the ground for a more thorough examination. So, this technology can be used to screen the population.

# CONCLUSION

This work describes a small part of medical, psychological and physiological applications of "MultiNeuron" series neural simulators. A series of diagnostic and predicting systems made on its basis are on probation in clinical practice.

Research applications of neural simulators in medicine are not less important than clinical ones. Medical researchers find the neural simulators an interesting interlocutor and new questions arising in communication with the neural networks seem to be the most important gain.